\documentclass{iaus}
\usepackage{graphicx}

\title[Spiral Galaxies in the SAURON Survey] 
{Spiral Galaxies in the SAURON Survey}

\author[Peletier and the SAURON team]   
{Reynier F. Peletier$^1$,
Katia Ganda$^1$, Jes\'us Falc\'on-Barroso$^2$, \break 
Roland Bacon$^{3}$, Michele Cappellari$^{4}$, Roger L. Davies$^{4}$, \break 
P. Tim de Zeeuw$^{5}$, Eric Emsellem$^{3}$, Davor Krajnovi\'c$^{4}$, \break
Harald Kuntschner$^{6}$, Richard M. McDermid$^{5}$, 
Marc Sarzi$^{7}$, \and Glenn van de Ven$^{8}$}

\affiliation{$^1$Kapteyn Astronomical Institute, University of Groningen,
NL-9700 AV Groningen \break email:
peletier@astro.rug.nl \\[\affilskip]
$^2$European Space Agency / ESTEC, Keplerlaan 1, NL-2200 AG Noordwijk, \break
$^3$Observatoire de Lyon, 9 av. Charles Andr\'e,
F-69230 Saint-Genis Laval,\\
$^4$Sub-Department of Astrophysics, University of Oxford, Oxford OX1 3RH, UK,
\break
$^5$Sterrewacht Leiden, University of Leiden, NL-2333~CA Leiden, \break
$^6$ST-ECF, European Southern Observatory, D-85748 Garching bei M\"unchen,
\break
$^7$Centre for Astrophysics Research, University of Hertfordshire, Hatfield, UK,
\break
$^8$Institute for Advanced Study, Einstein Drive, Princeton, NJ 08540, USA
} 

\pubyear{2007}
\volume{245}  
\pagerange{1-6}
\begin{document}

\maketitle

\begin{abstract}
We discuss some recent integral field spectroscopy using the SAURON instrument
of a sample consisting of 24 early-type spirals, part of the SAURON Survey, and
18 late-type spirals. Using 2-dimensional maps of their stellar radial velocity,
velocity dispersion, and absorption line strength, it is now much easier to
understand the nature of nearby galactic bulges. We discuss a few highlights of
this work, and point out some new ideas about the formation of galactic bulges.

\end{abstract}

\firstsection 
\section{Introduction}

One of the remarkable features of an image of a spiral galaxy is its central
light concentration. The recent Spitzer Space Telescope images once again showed
some beautiful examples of this, in particular the Andromeda galaxy (SSC
2006-14), the Sombrero (SSC 2005-11), M 51 (SSC 2004-19) and NGC 7331 (2004-12).
On the images one sees that every galaxy has a central light concentration due
to stellar light with relatively few features due to dust and star formation.
This light concentration is known as a galactic bulge. In the old days bulges
were known as old, metal rich, galaxy components (Whitford 1978). Nowadays,
there are strong indications that there is at least a large population of
galactic bulges which also contain younger stars. Details about this can be
found in the long review by Kormendy \& Kennicutt (2004).
Since research on bulges is making fast progress, however, I would like to
revisit some of the issues raised in this paper in the light of new 
observations that have been carried out by our group using the SAURON
instrument (Bacon et al. 2001), on the 
4.2m WHT at La Palma. 

Although spiral galaxies have been studied much less than ellipticals, there is
a considerable literature about the kinematics of stars and gas in the central
regions of spirals, and also on the stellar populations in these objects.
However, new insight can be obtained from the IFU-spectroscopy from 
SAURON, since it provides two-dimensional maps of several astrophysical
quantities, which are much easier to interpret than
one-dimensional profiles. At the same time, stellar populations can be combined
{\it locally} with kinematics, and the sample of spirals can be compared easily 
with the
large sample of nearby ellipticals observed with the same instrument. Since the
instrument covers a large field of view on the sky,
SAURON data are ideal to study the nature of galactic bulges, which, except for
the largest nearby galaxies, rarely dominate the disc outside the central
10$''$. At the same time one can learn something about differences between bulge
and inner disk.

In this proceedings we will discuss two samples of nearby galaxies: a sample of
24 S0/a-Sb galaxies, uniformly sampled in the luminosity - disk ellipticity
plane, part of the SAURON Survey (de Zeeuw et al. 2002), and a sample of 18
late-type spirals (Sb-Sd), optically selected with radial velocity smaller than
3000 km/s.  We will present a few highlights of a number of papers that we
have published recently: the kinematics of gas and stars and the emission line
strengths in the early-type spirals in Falc\'on-Barroso et al. (2006), the
absorption line strengths of the same sample in Peletier et al. (2007), the
stellar and gaseous kinematics and emission line strengths of the late-type
sample in Ganda et al. (2006), and the absorption line strengths of the same
sample in Ganda et al. (2007).

\section{SAURON Observations of Spiral Galaxies}

All spirals discussed here were observed using a single SAURON field of 33$''$
$\times$ 41$''$, sampled with lenslets of 0.94$''$ $\times$ 0.94$''$. Every
lenslet provides a spectrum with a wavelength range from 4790 to 5300 \AA, with
a spectral resolution of 4.2\AA\ FWHM. This spectral region contains several
absorption lines useful for studying stellar populations, as well as the [OIII]
emission line at 5007\AA. For the spiral galaxies it is very important to
separate the emission lines from the underlying absorption line spectrum. To do
this, the spectra were fitted with a linear  combination of stellar population
models of Vazdekis (1999), and gaussian emission lines (for details about this
procedure see Sarzi et al. 2006 and Falc\'on-Barroso et al. 2006). From the
cleaned spectra we  obtained the line indices H$\beta$, Mg\,$b$ and Fe 5015. 

In the way we described in Kuntschner et al. (2006) we determined ages,
metallicities and abundance ratios at every position, assuming that the stellar
populations there could be represented by a single-age, single metallicity
stellar population. In practise, we determined the SSP for which the line
strengths Fe 5015 , H$\beta$ and Mg\,$b$ fitted best in the $\chi^2$ sense
(Fig.~1).  Although we know that it is a great over-simplification to represent
the stellar populations (even locally) of a galaxy by a SSP, in some, especially
elliptical galaxies this approach gives a good first-order approximation.

\section{Stellar Populations in the Central Regions of Spiral Galaxies}

As can be nicely seen in the unsharp masked HST images, almost all spirals are
dusty in their central regions. This explains why it has been difficult to use
colours as tracers of stellar populations. An exception might be the work of
Peletier et al. (1999), who investigate colours of bulges for highly inclined
galaxy on the dust-free side. The line strength maps show that young stellar
populations are generally confined to a flat disk, with the young populations
distributed in a compact nuclear region, a ring, or a large central area, the
latter mostly in the faintest early-types or the late-type spirals. When
analysing the central line strengths in the line strength -- velocity dispersion
diagram, one finds that the early-type spirals can be both young and old, while
the late-type spirals clearly show a mix of stars of different ages. 

Here we would like to concentrate on two aspects, the behaviour of line strength
indices as a function of morphological type, and the central Mg/Fe ratio in
spirals. In Figure 1 we show the three indices H$\beta$, Mg~b and Fe 5015 as a
function of morphological types, also plotting the SAURON sample of E and S0
galaxies (Kuntschner et al. (2006)). The lower envelope of the H$\beta$ -- type
relation is filled with the oldest galaxies, while high H$\beta$ values are
found when a central burst of young stellar populations is present. In the two
other diagrams the old stellar populations form an upper envelope, with young
populations pushing the galaxies down. One can see that elliptical galaxies have
rather uniform stellar populations, with little scatter. For S0 galaxies the
same behaviour is seen, although there are a few outliers with strong central
star bursts. For the Sa galaxies, we see a large range of central line indices,
with the range decreasing towards the Sd galaxies. The figure shows that central
star formation is star burst-like in early-type spirals, and is much more quiescent
in late-type spirals. This agrees with the behaviour that is found when
analysing H$\alpha$ emission in spiral galaxies (Kennicutt 1998). It also shows
that there is a continuity in central stellar population properties when going
from early-type galaxies to the latest-type spiral galaxies, where some S0
galaxies  are very similar to early-type spirals.

\begin{figure}
\begin{center} 
\includegraphics[scale=0.75]{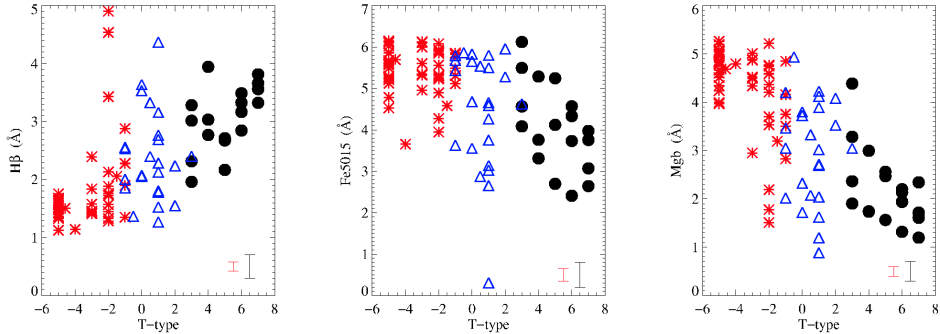}
\end{center} 
\caption[]{Several central line strength indices as a function of  morphological
type. Red asterisks are ellipticals and S0s from Kuntschner et al. (2006), blue
triangles early-type SAURON survey spirals and black dots late-type spiral
galaxies. 
} \end{figure}

Figure~2 shows Fe 5015 as a function of Mg~b. It shows that the central regions
of elliptical and S0 galaxies have Mg/Fe ratios that are larger than solar
(see Kuntschner et al. 2006 for more details), but also that especially
late-type spirals lie on the SSP models with solar Mg/Fe ratios. Although our
late-type spirals have stars with a considerable range in age and metallicity,
it would still be very hard to have models for which Mg/Fe is larger than solar
that agree with these observations. In fact, one expects late-type spirals, such
as our Milky Way, to have solar Mg/Fe ratios. They indicate slow star formation,
similar to what is happening in the solar neighbourhood. Both the bulge region
and the inner disk have solar Mg/Fe ratios. One sees that the
early-type spirals show a behaviour in between that of ellipticals and of
late-type spirals, with more scatter.

\begin{figure}
\begin{center} 
\includegraphics[scale=0.8]{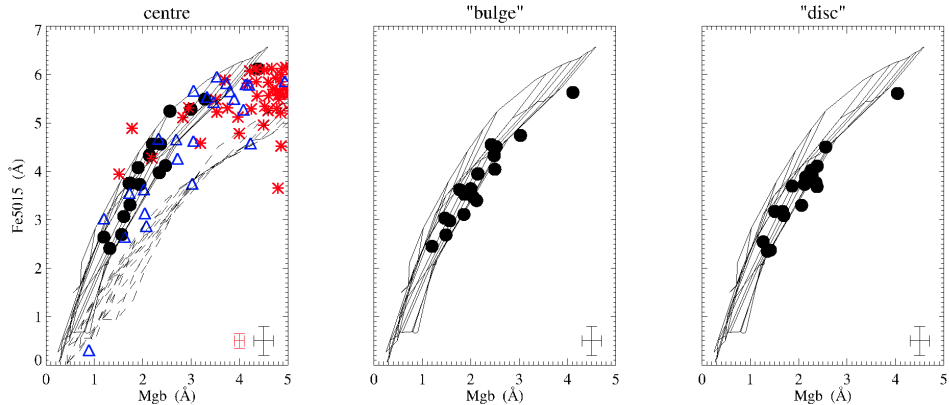}
\end{center} 
\caption[]{Index-index plots of Fe 5015 against Mg b, showing the [Mg/Fe]
abundance ratios in the centres of the galaxies. Symbols as in Figure 1. Shown
are model grids for single age-metallicity stellar populations, from Thomas et
al. (2003), for [Mg/Fe]=0 (full lines) and [Mg/Fe]=0.5 (dashed lines). One sees
that [Mg/Fe] for the late-type spirals is consistent with solar values. Plotted
are the central line indices on the left, the bulge in the middle, and the disk,
as covered by the SAURON field, on the right.}
\end{figure}

\section{Stellar Kinematics in Bulges - Sigma Drops}

The (v/$\sigma$) -- $\epsilon$ diagram has been a very powerful tool to
characterise the nature of galactic bulges. Kormendy \& Illingworth (1982)
showed that bulges fall close to the oblate rotator line in this diagram, the
line of an isotropic object flattened by its rotation. They, and also Davies et
al. (1983), showed that kinematically, these bulges behave the same as
intermediate mass ellipticals. The bulges for which kinematics was available at
that time were all large bulges of early-type spirals. Later, Kormendy could add
other bulges to the diagram, and found that some had larger (v/$\sigma$) than
expected for an isotropic rotator, showing more disk-like kinematics, and 
considerably complicating our understanding of galactic bulges.

At present, we know about many galaxies in which a thin, rotating disk is
dominating the light in the very inner parts, accompanied by a local minimum in
the velocity dispersion. The first observed cases of central velocity dispersion
minima date back to the late 80s and early 90s (e.g., Bottema 1989, 1993).
Several others were found from long-slit data (Emsellem et al. 2001, M\'arquez
et al. 2003). Now, SAURON observations show that 13 out of 24 Sa and Sab galaxies
showed a central local minimum in the velocity dispersion (F\'alcon-Barroso et
al. 2006). Also, half the late-type spirals have local velocity dispersion
minima in their centre, especially for the very latest types (Ganda et al.
2006).
We see a weak trend that galaxies with sigma-drops are younger than
those without (Peletier et al. 2007). The sigma-drops are probably due to
central disks that formed from gas falling into the central regions through a
secular evolution process. Simulations show that the disks will remain in place
for a long time after they have been formed (Wozniak \& Champavert 2006),
although they will slowly heat up with time. Such central disks are also known
among elliptical and S0 galaxies (e.g. NGC 4526, Emsellem et al. 2004, NGC 7332,
Falc\'on-Barroso et al. 2004), but rare.

\section{Our New Picture of Galactic Bulges}

From our study of the stellar populations and kinematics of the central regions
of spiral galaxies we infer that galactic bulges have more than one physical
component: generally they contain a slowly-rotating, elliptical-like component,
and one or more fast-rotating components in the plane of the galaxy, all
co-existing in the same galaxies. This picture also nicely explains the fact
that bulge populations in general are very similar to those in the disc (e.g.
Peletier \& Balcells 1996). 
In more than half of the SAURON early-type spirals sigma-drops occur. These
correspond to central disks, which sometimes  contribute most of the light
corresponding to the bulge resulting from the photometric  decomposition. HST
images shows that the central disks correspond to dusty regions,  often showing
spiral structure. They  also often, but not always, contain younger stellar
populations than the regions outside the  central disks.

We illustrate our picture using 2 galaxies (Fig.~3):  The first one, NGC 4274,
for which the surface brightness profile is fitted best by a S\'ersic profile
with $n$=1.3, has a strong sigma-drop. In the region where it dominates the
stars rotate fast and have a low velocity dispersion (SAURON). Such an inner
disc
would be called a pseudo-bulge by Kormendy \& Kennicutt (2004). It is
dusty, has spiral structure (HST image) and shows younger stellar populations
(SAURON). Note also that as one goes out on the minor axis, the stellar
distribution becomes smooth, and the line strengths show values corresponding to
old stellar populations. This region corresponds to an elliptical-like bulge, to
which Kormendy \& Kennicutt (2004) would refer as a classical bulge. NGC 4274
therefore would contain a pseudo-bulge \textit{AND} a classical bulge. Making the
classical bulge a bit larger, and the pseudo-bulge smaller, one gets an object
such as NGC 3623 (bottom of Fig.~3). Here the spheroidal bulge dominates the
light, apart from the very inner regions. A sigma-drop is seen, but weaker than
in NGC 4274 (Falc\'on-Barroso et al. 2006). The S\'ersic fit for the bulge gives
$n$=3.4. Dust is associated with the central disks, but the SAURON maps show
that the stellar populations in it are not different from the bulge
outside it. The comparison between NGC 3623 and NGC 4274 shows that both objects
are very similar, but that the inner disk to elliptical bulge ratio in both
galaxies is different.

To conclude, we would like to point the reader to the similarities between
elliptical and spiral galaxies. Not only can elliptical galaxies consist of
several components (inner disks, KDCs etc.), the same can be the case for spiral
galaxies, with their large, thin disk, thick disk, hot bulge, inner disk etc.

\begin{figure}
\begin{center} 
\includegraphics[scale=0.8]{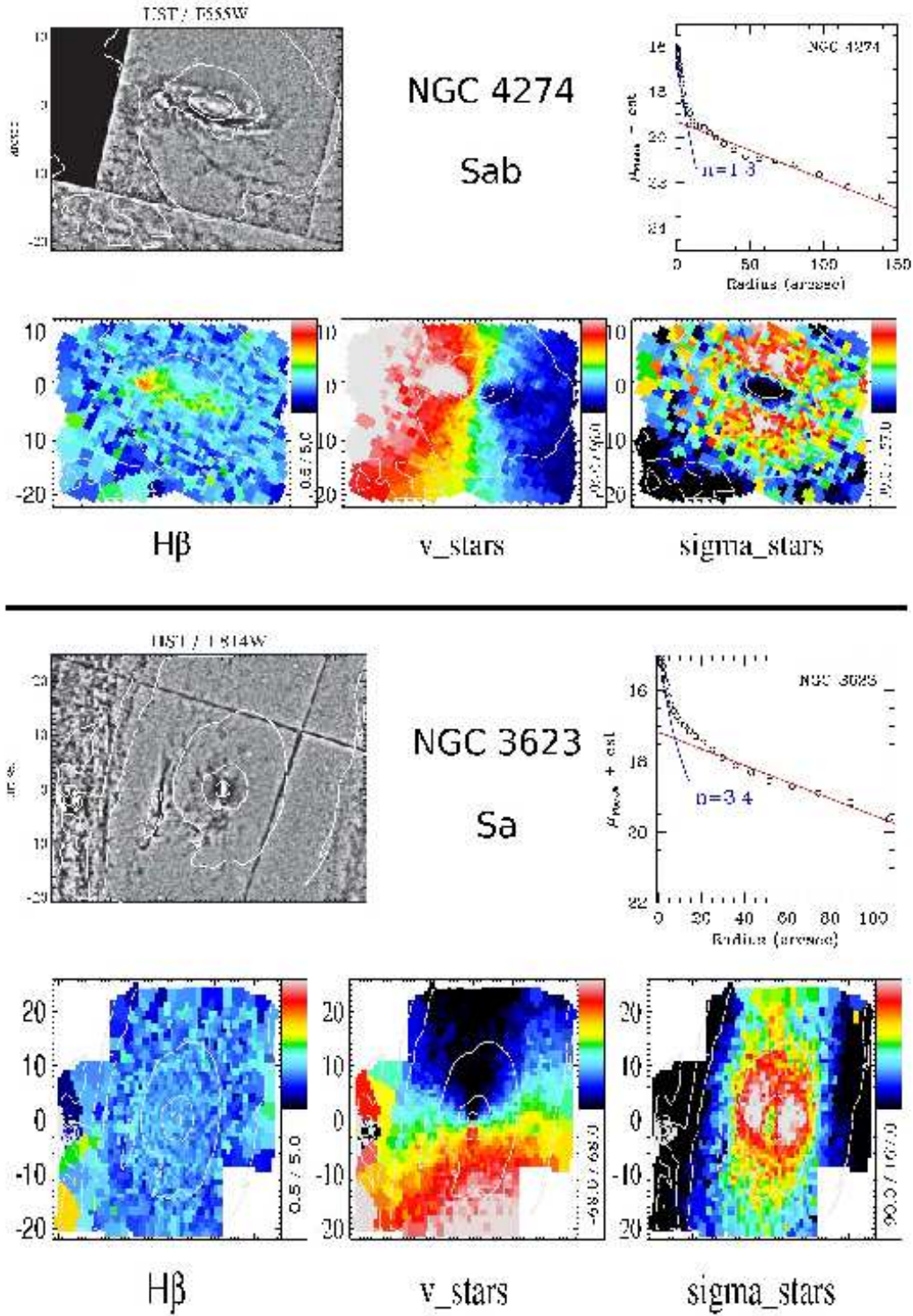}
\end{center} 
\caption[]{Diagnostic observations of NGC 4274 and NGC 3623, two early-type
spirals. For each galaxy is shown in the top left: unsharp-masked F555W HST
image, showing the places of non-negligible extinction (from Falc\'on-Barroso et
al. 2006). Top right: major axis surface brightness profile, from the same HST
data. A bulge-disk decomposition is also shown, with an exponential disk and a
S\'ersic bulge. Bottom row (from left to right): H$\beta$ absorption line map
(Peletier et al. 2007),  stellar velocity and velocity dispersion map
(Falc\'on-Barroso et al. 2006). Overlayed on all maps is the reconstructed
SAURON intensity.}
\end{figure}

\end{document}